\documentclass[preprint,aps]{revtex4} 

\usepackage{graphicx}

\begin{document}
\title{The Montevideo interpretation of quantum mechanics: frequently
asked questions}

\author{Rodolfo Gambini$^{1}$,
and Jorge Pullin$^{2}$} 
\affiliation {1. Instituto de F\'{\i}sica,
Facultad de Ciencias, Igu\'a 4225, esq. Mataojo, Montevideo, Uruguay. \\
2. Department of Physics and Astronomy, Louisiana State
University, Baton Rouge, LA 70803-4001} 

\date{March 12th 2009}

\begin{abstract}
  In a series of recent papers we have introduced a new interpretation
  of quantum mechanics, which for brevity we will call the Montevideo
  interpretation. In it, the quantum to classical transition is
  achieved via a phenomenon called ``undecidability'' which stems from
  environmental decoherence supplemented with a fundamental mechanism
  of loss of coherence due to gravity. Due to the fact that the
  interpretation grew from several results that are dispersed in the
  literature, we put together this straightforward-to-read article
  addressing some of the main points that may confuse readers.
\end{abstract}
\maketitle

\parindent=0pt

In a series of recent papers we have introduced a new interpretation
of quantum mechanics, which for brevity we will call the Montevideo
interpretation. The most complete description published up to date is
the essay \cite{fqxiessay}. Here we clarify some points about the
interpretation in the form of a set of questions and answers.

\begin{itemize}

\item{\bf Briefly, what is the Montevideo interpretation of quantum 
mechanics?}

This interpretation explains the emergence of the classical world via
decoherence through the interaction with the environment plus loss of
coherence of the quantum theory when studied with real clocks and
measuring rods (a pedagogical discussion of why there is loss of
coherence when one considers real rods and clocks is in
\cite{obregon}). The combined effect of both losses of coherence
implies that all information about quantum coherence in the system
plus apparatus plus environment becomes inaccessible.

After a while, there is no experimental arrangement that is able to
decide if the evolution of the state of the complete quantum system
was unitary or suffered a collapse. Whenever such situation of {\em
undecidability} is reached, the interpretation assumes that an event
(measurement) takes place \cite{fqxiessay}.

\item{\bf Is the interpretation realist?}
  
  This interpretation is realist in the sense that it attempts to
  yield a description of reality based on available physical theories.
  We assume that physical reality is constituted by two fundamental
  elements: the states and the events. The latter include observable
  events, (the observable phenomena). From the latter we define the
  physical properties of the system, which are all quantitative
  expressions of phenomena.  The states characterize the disposition
  of the system to produce events with certain probabilities. We admit
  that certain elements of reality are undecidable, that is, the
  laws of physics may not describe reality completely (see below).

\item{\bf What is this loss of coherence due to real clocks and rods?}

Ordinary quantum theory evolves unitarily in terms of a time parameter
$t$.  Such parameter is assumed to be a completely classical variable,
which we know with arbitrary accuracy. In reality, no variable in
nature is purely classical, so when we use a real clock, we are not
measuring $t$ but a new variable $T$. Even if one chooses a very
classical clock, trying to correlate $T$ with $t$ as well as possible,
such a variable has some level of quantum fluctuations and other
inaccuracies. Our inability to measure time accurately implies that pure
states evolve into mixed states, the superposition being that ``a
state at time $T$'' really corresponds, due to the clock's
inaccuracies, to a probabilistic distribution of several possible
values of $t$ upon evolution.  The spread of the distribution grows
as the system evolves. Similar effects arise in quantum field
theories when one measures the spatial coordinates. A pedagogical
introduction can be seen in \cite{obregon}.

\item{\bf Couldn't one eliminate the effects by using better clocks?}

At some point one will encounter fundamental limitations to the
accuracy of a clock, so the effect will always be there. Some authors
have argued that the fundamental limits are surprisingly large (in
Planck units), by noting that every measurement requires the
expenditure of energy and if you expend too much energy trying to be
accurate your clock becomes a black hole. This is the rough argument,
there are variations on it in the literature and they all seem to come
to remarkably similar limits
\cite{ng}.  This particular  limit states that the inaccuracy of a
clock that measures a time $T$ is given by $\delta T \sim T_{\rm
Planck}^{2/3} T^{1/3}$, where $T_{\rm Planck}\sim 10^{-44}s$. One sees
that for laboratory typical times of a few hours, $\delta T\sim
10^{16} T_{\rm Planck}\sim 10^{-28}s$.

\item{\bf But aren't those fundamental limits quite questioned in the 
literature?}

Indeed (see \cite{aleman}). We have, however, carried out detailed
calculations \cite{pintosrisso} with much more conservative estimates
on the clock inaccuracies, for instance that the fluctuations in the
clock grow with any power of the time measured ---even close to
zero---. We find that the effects we are discussing are still present
even for systems having a number of degrees of freedom several orders
of magnitude smaller than what would consider ``the classical world''.

\item{\bf Does one need to involve quantum gravity for all this?}

Not really. It is true that in quantum gravity the theory is generally
covariant and one naturally accepts that there is no absolute time $t$
(the so-called ``problem of time'' \cite{kuchar}). A favored solution
to this problem is to build the theory relationally by choosing a
clock from among the physical variables \cite{time} (for readers with
more interest in how the fundamental decoherence relates to the
problem of time in quantum gravity, see the next to last
question). Gravity also plays a role in setting a fundamental limit to
how accurate a clock can be. Up to now detailed models of this from
quantum gravity are lacking, the bounds placed are phenomenological
semi-classical estimates. That in part justifies why they are
criticized in the literature.

\item{\bf But ultimately, you are just talking about loss of coherence.
In addition to the degrees of freedom environment you now just added the 
degrees of freedom of the clock. How could that change things so much?}

What makes things different is that a clock is a very special set of
degrees of freedom (in practice it is also a quite large set of
degrees of freedom, at least if one wants it to remain accurate for a
while). These degrees of freedom are being used to establish a
property of the rest of the system. This is very different from the
degrees of freedom of the environment. A clock is also special in that
it is supposed to stay accurate during the lifetime of the experiment
one studies (and better yet, beyond).  Real clocks set limits on what
is experimentally accessible, in particular from a pure state after it
has evolved for a certain amount of time.

\item{\bf One of the objections to environmental decoherence is that
the evolution of the ``system plus measuring device plus environment''
viewed as a closed system is unitary and one could recover quantum
coherence eventually, by waiting a long time}

The loss of coherence due to the loss of the clocks' accuracy is such
that it gets worse the long one waits. So waiting longer does not
help. In fact it guarantees that the ``revivals'' of coherence several
authors have studied for closed macroscopic systems are killed off by the
loss of coherence due to the clock. A concrete calculation was worked
out in \cite{foundations,pintosrisso}.

\item{\bf d'Espagnat \cite{despagnat} has argued in models like Zurek's
\cite{zurek} that one could in principle measure global observables of
the system that would indicate if collapse has occurred, even without
the need of observing ``revivals''. Can't the observables be used 
to decide if the system collapsed or not?}

It is obviously difficult to attempt to give a general answer to such
a question since global observables require including quantum
measurements of properties of the environment, which are in practice
virtually impossible to carry out.  We have tried to model the
situation in a modification of Zurek's example where one has some
control of the environment.  This allows a chance at least in
principle of measuring the global observable proposed by
d'Espagnat. We have found that fundamental limitations of quantum
mechanics and the use of real clocks disallow its measurement
\cite{pintosrisso}. Notice that, at least in this example, these
limitations are not practical in nature, but fundamental, resulting
from applying the laws of quantum mechanics.

\item{\bf But there are other problems with the solution to the
measurement problem by decoherence. What about the and/or problem as
Bell \cite{bell} called it, or ``the problem of outcomes''?}.

The ``problem of outcomes'' is that although decoherence may have
taken place, taking the system plus measurement apparatus to a state
in which their density matrix is quasi-diagonal, the system remains in
a superposition of states, each of them associated with different
macroscopic behaviors. States corresponding to the diagonal
of the density matrix, coexist with each other. How does one go from
this situation to the classical world, where one of those alternatives
and one only is realized? How does one go from a situation where one
has alternatives ``A {\em and} B'' to ``A {\em or} B''?  In our
approach we argue that there exist situations where the decoherence
due to the environment and the use of real rods and clocks implies
that ``A {\em} and B'' provided by a unitary evolution and the ``A
{\em or} B'' result we would attribute to a reduction, are
indistinguishable. When one reaches this state of {\em
undecidability}, an event has taken place, without affecting any
experimental outcome.

\item{\bf  What happens to the quantum state, does it collapse or not?}

Since we claim an event (measurement) has happened when one 
cannot decide if the state has collapsed or has evolved unitarily, one
clearly cannot decide if it collapsed or not. We have proposed
that a point of view is  that both alternatives be accepted as possible
for the system. Entering the realm of philosophy, this means we are
taking a ``regularist'' point of view. In philosophy there are different attitudes
that have been taken towards the physical laws of nature (see for
instance
\cite{stanford}).  One of them is the ``regularity theory'', many times 
attributed to Hume \cite{hume}; in it, the laws of physics are
statements about uniformities or regularities of the world and
therefore are just ``convenient descriptions'' of the world.  Ernest
Nagel in {\em The Structure of Science} \cite{nagel} describes this
position in the following terms: {\em ``...according to Hume [physical
laws consist] in certain habits of expectation that have been
developed as a consequence of the uniform but de facto conjunctions of
[properties]."}  The laws of physics are dictated by our experience of
a preexisting world and are a representation of our ability to
describe the world but they do not exhaust the content of the physical
world. A second point of view sometimes taken is the ``necessitarian
theory''
\cite{stanford}, which states that laws of nature are ``principles''
which govern the natural phenomena, that is, the world ``necessarily
obeys'' the laws of nature. The laws are the cornerstone of the
physical world and nothing exists without a law. The presence of the
undecidability we point out suggests strongly that the ``regularity
theory'' point of view is more satisfactory since the laws do not
dictate entirely the behavior of nature. Notice that the freedom of 
the system to collapse or not is not governed by any probabilistic
rule. This introduces a notion of free will into our description
of nature \cite{fqxiessay}.



\item{\bf  Does your point of view argue against the many-worlds
interpretation?}

Not necessarily. The many-worlds interpretation is compatible with
what we are claiming. However, it becomes less compelling. The
many-worlds interpretation was introduced in part to address the
``and/or problem'' by assigning different worlds to different
alternatives. Since we are given a criterion for the passage from and
to or, the other alternatives (and other worlds) are therefore not
really necessary. That does not mean that they are precluded by our
point of view.

\item{\bf  Does your point of view argue against the ``modal'' interpretations?}

Well, let's see. Dieks \cite{dieks} lists the following set of properties as
shared by all modal interpretations:

$\diamond$ The interpretation is based on the standard formalism of
quantum mechanics. [OK, plus quantum gravity].

$\diamond$ The interpretation is realist, that is, it aims at
describing how reality would be if quantum mechanics were true. [OK].

$\diamond$ Quantum mechanics is a fundamental theory, which must
describe not only elementary particles but also macroscopic objects.
[OK].

$\diamond$ Quantum mechanics describes single systems: the quantum
state refers to a single system, not to an ensemble of systems. [OK].

$\diamond$ The quantum state of the system (pure state or mixture)
describes the possible properties of the system and their
corresponding probabilities, and not the actual properties. The
relationship between the quantum state and the actual properties of
the system is probabilistic. [OK].

$\diamond$ Systems possess actual
properties at all times, whether or not a measurement is performed on
them.  [NO, actual properties only appear when the system becomes
undecidable].

$\diamond$ A quantum measurement is an ordinary physical
interaction. There is no collapse: the quantum state always evolves
unitarily according to the Schroedinger equation.  [NO, this cannot be
decided]

$\diamond$ The
Schroedinger equation gives the time evolution of probabilities, not of
actual properties. [OK].

As we see, there are a lot of points in common, but not all properties
of a ``modal'' interpretation are satisfied by ours. In modal
interpretations a central problem is the determination of when events
occur. There have been many proposals for such a point and there has
not emerged a unanimous consensus on this (see \cite{castagnino} and
references therein)

\item{\bf  A friend of mine works in string theory and claims the theory is
unitary and your effect therefore is wrong.}

String theory has been studied in many situations and in all cases
where it makes sense to talk about an evolution as a function of a 
parameter, such evolution is unitary.  
In these situations however, we are just like in
ordinary quantum mechanics, when formulated in terms of a classical
parameter $t$ (ok, the story is slightly more complicated since we are
talking about a field theory but we mentioned the effect is present
there too, supplemented with additional spatial decoherence
effects). Our point is that such classical parameters are not accessible in
reality. So if one were to formulate string theory in terms of a real
clock $T$ one would have the same types of effects we discuss
here. Ask your friend if she believes that string theory solves the
``problem of time'' of quantum gravity.  The reply will likely be that
the problem persists, it is just that string theorists tend to
concentrate on other issues and situations, where the problem of time
is not relevant.

\item{\bf  This loss of coherence of yours could be made arbitrarily large
by choosing a very inaccurate clock. Has this been observed 
experimentally?}

Actually yes, although no experimentalist has deliberately used a bad
clock to observe the loss of coherence, certain correlations in 
experiments involving Rabi oscillations can be interpreted as using
some of the atoms as ``bad clocks'' and indeed one sees loss of 
coherence like the one we mention \cite{bonifacio}.

%

\item{\bf Will the fundamental loss of coherence arise naturally in a
technical treatment of the problem of time in quantum gravity, or 
is it a choice?}

It does. In general relativity there is no notion of absolute time. In fact,
there is no absolute notion. All physical predictions have to be
formulated as relations between physical quantities.  This has been
recognized since the early days of general relativity through
Einstein's ``hole argument'' \cite{hole}. In particular the notion of
time has to emerge ``relationally''. One possible way of attacking this
was introduced by Page and Wootters \cite{PaWo}.  In their
proposal one takes any physical quantity one is interested in studying
and chooses another physical variable that will act as ``clock''. One
then studies how the first variable ``evolves as a function of the
second one''. In this view, time does not play any preferred role among
other physical variables. This is in contrast to 
ordinary quantum mechanics where
one has to unnaturally assume that time is supposed to be the only
variable in the universe not subject to quantum fluctuations. In spite
of the simplicity and naturalness of this proposal to tackle the
problem of time in quantum gravity, technical problems arise. The
problems are related with what one considers to be physical quantities
in a theory like general relativity. Usual things that one may
consider physical quantities, like ``the scalar curvature of
space-time at a given point'' are not well defined objects in general
relativity.  The problem is what is ``a given point''? Points in space
have to be defined physically in general relativity. One can
characterize a point as a ``place where something physical happens''
(for instance, a set of physical fields takes certain values). Then
one could ask ``how much is the curvature at that point''. The end
result is indeed physical. But it is again a relation between the
values of curvatures and fields. Such relation is given and
immutable. How could one construct a clock out of something immutable?
It appears that the only things that are physical are immutable
relations and the only things that evolve are the members of the
relations, like the curvatures and fields. In technical terms, what
one can consider as physical observable in general relativity is a
quantity that is left invariant under the symmetries of the theory, or
in the canonical language, that commutes with the constraints. Since
one of the constraints is the Hamiltonian, physical quantities do not
evolve. Therefore they cannot work as clocks. This created problems
\cite{kuchar}
for the Page--Wootters proposal.  A way out was sought by trying to
establish relations not between physical quantities but between
mathematical quantities one uses to describe the theory that are not
directly measurable (like for instance, the components of the metric
at a point). Far from helping, this led to significant technical
problems since one ends formulating the theory in terms of
unobservable quantities. Ultimately it was shown in model systems that
the proposal cannot be used to compute elementary things, for instance
quantum probabilities of transition \cite{kuchar}.

The observation we have recently made \cite{time} is that the
Page--Wootters construction can be rescued by using Rovelli's proposal
of ``evolving constants of the motion'' \cite{rovelli}, a concept that
can be traced back to DeWitt, Bergmann and Einstein himself.  This
idea is to introduce genuinely observable physical quantities,
i.e. relations between magnitudes as we highlighted above, but that
depend on a continuous parameter. If one imagines evolution as changes
in such a parameter, one can actually construct the relational
description of Page and Wootters and show that it actually leads to
the correct quantum probabilities of transition, at least in model
systems \cite{time}. The beauty of the complete construction is that
the continuous parameter in the evolving constants completely drops
out at the end of the day and the formulation remains entirely written
in terms of truly observable physical quantities, even in the extreme
situations that can develop in physics when quantum gravity effects
become important. The calculation of the propagator with this 
construction, at least in model systems, yields the usual propagator
at leading order, but one has corrections to it due to the mechanisms
of loss of coherence that we are referring to in this paper.

\item{\bf What is a good reference for all this?}
  
  We have an Fqxi essay where these ideas are presented for
  non-experts \cite{fqxiessay}. A pedagogical detailed discussion of
  the fundamental loss of coherence is in \cite{obregon}. Papers
  discussing specifically issues of the problem of measurement in
  quantum mechanics are \cite{foundations} and \cite{pintosrisso}.

\end{itemize}

Portions of this work were developed in collaboration with Luis
Garc\'{\i}a-Pintos, Rafael Porto, and Sebasti\'an Torterolo.  This
work was supported in part by grants NSF-PHY-0650715, FQXi, CCT-LSU,
Pedeciba, ANII PDT63/076 and the Horace Hearne Jr.  Institute for
Theoretical Physics.

\end{document}